\documentclass[useAMS]{mn2e}
\usepackage{epsfig}
\title[Globular cluster black hole variability]{Fading of the X-ray flux from the black hole in the NGC~4472 globular cluster RZ~2109}

\author[Maccarone et al.]{Thomas J. Maccarone\\ School of Physics and
Astronomy, University of Southampton, Hampshire SO17 1BJ,United
Kingdom\\ \newauthor Arunav Kundu\\ Eureka Scientific, 2452 Delmer
Street Suite 100, Oakland, CA 94602-3017, USA\\ \newauthor Stephen
E. Zepf\\ Department of Physics and Astronomy, Michigan State
University, East Lansing, MI 48824, USA\\ \newauthor Katherine
L. Rhode\\ Department of Astronomy, Indiana University,727 East 3rd
Street, Bloomington, IN 47405-7105, USA }

\begin{document}
\def\ltsim{\mathrel{\rlap{\lower 3pt\hbox{$\sim$}}
        \raise 2.0pt\hbox{$<$}}}
\def\gtsim{\mathrel{\rlap{\lower 3pt\hbox{$\sim$}}
        \raise 2.0pt\hbox{$>$}}}

\date{}

\pagerange{\pageref{firstpage}--\pageref{lastpage}} \pubyear{}

\maketitle

\label{firstpage}

\begin{abstract}
We present the results of new X-ray observations of
XMMU~122939.7+075333, the black hole (BH) in the globular cluster
RZ~2109 in the Virgo Cluster galaxy NGC~4472.  A combination of
non-detections and marginal detections in several recent {\it Swift}
and {\it Chandra} observations show that the source has varied by at
least a factor of 20 in the past 6 years, and that the variations seem
not just to be ``flickering.''  This variation could be explained with
changes in the absorption column intrinsic to the source no larger
than those which were previously seen near the peak of the 1989
outburst of the Galactic BH X-ray binary V404~Cyg.  The large
amplitude variations are also a natural expectation from a
hierarchical triple system with Kozai cycles -- the mechanism recently
proposed to produce BH-white dwarf (WD) binaries in globular clusters.
On the other hand, variation by such a large factor on timescales of
years, rather than centuries, is very difficult to reconcile with the
scenario in which the X-ray emission from XMMU~122939.7+075333 is due
to fallback of material from a tidally destroyed or detonated WD.
\end{abstract}


\section{Introduction}

The dense stellar systems in globular clusters can produce a variety
of classes of exotic binary stars.  Until recently, there was no
convincing evidence for any BH X-ray binaries in globular clusters.
All of the Milky Way's globular clusters which contain bright X-ray
binaries have shown Type I X-ray bursts, which are caused by
thermonuclear runaway on a solid surface, requiring a neutron star
(NS) accretor.  Two of the clusters contain pairs of X-ray binaries
which have become bright, although even in those cases, both objects
are believed to be NSs.  In NGC~6440, two distinct pulse periods have
been seen from the cluster (Altamirano et al. 2010), while in M~15,
the source which is not thought to be a burster is likely to be a NS
based on Doppler tomography (van Zyl et al. 2004).

With the advent of the {\it Chandra} X-ray observatory, with its
excellent angular resolution and sensitivity, it became possible to
associate X-ray sources with luminosities in excess of $10^{39}$
ergs/sec with globular clusters (e.g. Sarazin et al. 2000; Angelini et
al 2001).  However, such luminosities can potentially be produced by
multiple bright NSs in the same cluster; strong variability is one of
the few ways to distinguish between a single BH X-ray binary and many
bright NSs (Kalogera et al. 2004).  In recent years, several globular
cluster X-ray sources have shown variability by amounts greater than
the Eddington luminosity for a NS (Maccarone et al. 2007; Brassington
et al. 2010; Shih et al. 2010; Maccarone et al. 2010), providing the
first strong evidence for BHs in globular clusters.  These discoveries
have been particularly exciting in view of theoretical work suggesting
that stars much heavier than the mean stellar component in a cluster
(such as stellar mass BHs) should be efficiently ejected (Spitzer
1969).  One the other hand, more recent, more detailed numerical
calculations have found black hole retention fractions similar to
those for NSs (Mackey et al. 2007; Moody \& Sigurdsson 2009).

XMMU~122939.7+075333 in the globular cluster RZ~2109 (Rhode \& Zepf
2001) in NGC~4472 was the first ultraluminous globular cluster X-ray
source to show strong X-ray variability (Maccarone et al. 2007).
Additionally, this source shows strong, broad [O~III] emission lines
[Zepf et al. 2007;2008 (Z08)].  The emission lines are sufficiently
strong and broad that they cannot be produced through virial motions
around a BH less than about $3\times10^4M_\odot$ (Z08; Porter 2010),
so the system is much more likely to be an accreting stellar mass BH
with a strong disk wind than an intermediate mass BH (Z08), since
such winds are likely only near the Eddington luminosity
(e.g. Proga 2007).

One way to account for the large ratio of oxygen relative to other
species in the optical spectrum is through a WD or some other highly
evolved donor star (Gnedin et al. 2009).The high luminosity then
implies that the donor star is likely to be a WD in a short orbital
period binary.  Recently, Ivanova et al. (2010) have suggested that
the most efficient way to produce such a system is to make it the
inner binary system in a hierarchical triple star system, with the
eccentricity induced by the Kozai (1962) cycles grinding down the
orbit into contact on a short timescale.  In this Letter, we report on
the discovery of large amplitude variability of this source,
consistent with it turning off as an X-ray source, and discuss it in
terms of eccentricity-induced accretion rate cycles.  We also consider
the possibility that the source varies because of changes in the
foreground absorption.

\section{Data}
\subsection{Observations already in the literature}
Several previous observations have been made with sufficiently good
angular resolution to allow for measurements of the flux of
XMMU~122939.7+075333.  The dates and luminosities of the published and
new observations of XMMU~122939.7+075333 are presented in Table
\ref{luminosities}.  Most of the past observations have been
previously summarized by Shih et al. (2008), but we repeat the results
here for completeness.  ROSAT observed NGC~4472 with its High
Resolution Imager for 27000 seconds of live time spread over June and
July of 1994.  XMMU~122939.7+075333 was reported as IXO 60 in the
``intermediate X-ray object'' catalog of Colbert \& Ptak (2002).  They
estimated a source luminosity of $10^{39.9}$ ergs/sec from 2-10 keV,
assuming a $\Gamma=1.7$ power law.  Due to the poor spectral
resolution of the HRI combined with the fact that the measurements
were actually made from 0.1-2.4 keV, there is considerable uncertainty
in the counts-to-energy conversion factor.  Re-evaluating the flux for
XMMU~122939.7+075333 using a softer spectral model can reduce the
inferred luminosity dramatically; we have taken the observed count
rate of $4\times10^{-3}$ counts/sec and computed fluxes using the
W3PIMMS tool and have found that for a 0.2 keV blackbody, the inferred
luminosity is 3$\times10^{39}$ ergs/sec.  We find it safe to say only
that the luminosity during this epoch is likely to be greater than
$10^{39}$ and less than $10^{40}$ ergs/sec.
\begin{table}
\begin{tabular}{lll}
\hline Observatory&Date&Luminosity\\ \hline {\it Einstein}&July 1979&
$\ltsim$$10^{40}$ergs/sec \\ {\it ROSAT}&June-July 1994&
$\sim5\times10^{39}$ ergs/sec\\ {\it Chandra}&12 June 2000&
$5\times10^{39}$ ergs/sec\\ {\it XMM-Newton}& 5 June 2002&
$4\times10^{39}$ ergs/sec\\ {\it XMM-Newton}& 1 January 2004&
$4\times10^{39}$ ergs/sec\\ {\it Swift} & 2007 Dec./2008 Jan. & $<3\times10^{39}$ergs/sec\\ {\it Chandra}& 27
February 2010& $\approx$$1\times10^{38}$ erc/sec\\
{\it Swift} & late March 2010& $<1.5\times10^{39}$ ergs/sec\\
\end{tabular}
\caption{The X-ray luminosities of XMMU~J122939.7+075333 from
long-look observations.  The ROSAT observation has a factor of $\sim2$
uncertainty in the luminosity due to uncertain count-to-energy
conversion.  The 2004 XMM observation entry is the source luminosity
in the bright epoch.  It is also the source luminosity in the faint
epoch under the assumption that its variability within the observation
is due to a change in absorption instrinsic to the source. The source
luminosity in the faint epoch if one accounts only for the Galactic
foreground absorption is about a factor of 3 lower.  The 2010 {\it
 Chandra} detection is statistically marginal.  The {\it Swift} upper
limits are given at the 95\% confidence level.}
\label{luminosities}
\end{table}

XMMU~122939.7+075333 was observed by Chandra on 12 June 2000, showing
a soft spectrum (well-fitted by a $kT_{in}=0.2$ keV disk blackbody
model) with a luminosity of about $5\times10^{39}$ ergs/sec (Shih et
al. 2008); by XMM-Newton on 5 June 2002 with a luminosity of
$4\times10^{39}$ found from the 2XMM survey (Watson et al. 2009).
Finally, it was observed by XMM on 1 January 2004, where the rapid
variability was found.  In the 2004 XMM observation, the source was at
$4\times10^{39}$ for about 10 kiloseconds, then dropped by a factor of
7 in count rate, with the drop consistent with a change in the
foreground absorption column, with no change in the instrinsic source
spectrum (Maccarone et al. 2007).  The spectrum of the faint part of
the XMM observation would have produced $\sim$200 counts in the 2010
{\it Chandra} observations -- the variability is clearly strong and
significant.

We note additionally that {\it Einstein} observed NGC~4472 in 1978, and did
not make a detection.  The noise level in the data set for the
off-axis detected sources was $2\times10^{-3}$ cts/sec (Harris et al
1993), so the non-detection implies a count rate below about
1$\times10^{-2}$ cts/sec.  Using W3PIMMS to convert this count rate to
a luminosity, we find an upper limit of $1.5\times10^{40}$ ergs/sec
assuming a $\Gamma=1.7$ power law, and about half that using blackbody
models in the range from $kT_{BB}=0.2-0.5$ keV in temperature,
assuming a distance of 16 Mpc (Macri et al. 1999).

\subsection{New Chandra observations}
We observed NGC~4472 with {\it Chandra} on 27 February 2010.  The
primary motivation for that observation was to look for variability
from bright sources in the inner regions of the galaxy, so
XMMU~122939.7+075333 lies on the edge of the ACIS-S3 chip.  We
produced an exposure map to determine the effective exposure time at
the source position and found it to be about 20000 seconds.

We have run WAVDETECT on the filtered events list, using the standard
$10^{-6}$ null hypothesis probability which is set to ensure that
there will typically be $\sim1$ chance detection over an ACIS chip.
No source is detected at the position of RZ~2109.  Given the
vignetting and dither pattern's effects, the effective exposure time
at the source's position, as estimated from the exposure map, is about
20000 seconds.  We then use aperture photometry with radius 12.2
pixels, appropriate for containing all the flux from the source at
this angle
off-axis\footnote{http://cxc.harvard.edu/ccw/proceedings/03\_proc/presentations/allen/index.html},
and we detect 19 counts between 0.5 and 8 keV.  The expected rate of
background photons over a region this size at this position is 10.8.
The net number of counts in the region is then $8.2\pm3.3$ -- giving a
roughly 1\% chance of a fluctation in the background producing this
flux at this position.  Using W3PIMMS with either a $\Gamma=1.7$ power
law or a $kT=1$ keV blackbody (spectra roughly consistent with the
low/hard and high/soft states for stellar mass black holes), we find a
flux of about $4\times10^{-15}$ ergs/sec/cm$^2$, which corresponds to
a source luminosity of $10^{38}$ ergs/sec.  The 3$\sigma$ upper limit
from the source is about $2\times10^{38}$ ergs/sec.

\subsection{Swift observations}
We have triggered the Swift X-ray Telescope to observe
XMMU~122939.7+075333 five times -- on 2007 December 25 and 2 January
2008, roughly simultaneously with our Keck spectroscopy (Zepf et
al. 2007;2008), and three times in late March of 2010 in response to
the non-detection in the {\it Chandra} observations presented above.
The observation ID numbers are 0031078001 through 0031078005.  We also
note that an additional short XMM observation was made at about the
same time as the Keck spectrum, but that strong flaring background
prevented those data from being useful.

We analyse the Swift data using the standard cleaned events files,
after final versions had been entered into the archive.  A quick
visual inspection of the data revealed that the source would be, at
best, marginally detected by Swift in our observations.  For that
reason, we use a 20'' aperture to extract a number of detected
photons.  This radius corresponds to the 75\% encircled energy region
for Swift XRT (need ref for Swift XRT PSF), but it allows for a much
lower background count rate than using the standard 90\% encircled
energy region of 47''.  We extract events in channels 50-500
(approximately 0.5-5 keV).  We use a 200'' aperture off-axis region to
estimate the background count rate.

We combine the two observations in 2007/2008 with one another to make
``Swift epoch 1'' and the three observations from 2010 to make ``Swift
epoch 2''.  For Swift epoch 1, we detect 4 source counts, and 95
background counts in a total of 3843 seconds.  Given that the
background region has a radius ten times as large as the source
region, we estimate that the source region should contain
$0.95\pm0.09$ background counts.  There is thus a 1.6\% chance that the
4 detected counts could be produced by Poisson fluctuations of the
background.  If we assume a spectrum of an 0.2 keV blackbody for the
source convolved with the foregound Galactic absorption and use
W3PIMMS to convert counts to energy, then the inferred source X-ray
luminosity is about $10^{39}$ ergs/sec for this marginal
detection.  A Poisson process with mean of 9.3 counts will produce 4
or fewer detected counts 5\% of the time.  We can then take as an
upper limit for the source count rate 8.3 counts over the time
interval, giving an upper limit to the source luminosity of about
$3\times10^{39}$ ergs/sec, with some additional uncertainty based on
the spectral model used to convert counts to energy.  This observation
thus provides weakly suggestive evidence that XMMU~122939.7+075333 had
already started to fade in the X-rays by late 2007.  The data from
Swift epoch 2 show 1 source photon and 134 background photons in 5003
seconds of summed exposure time -- the background count rate per pixel
is nominally higher than the source region rate.  The 95\% confidence
level upper limit on the number of source plus background counts is
about 5.8 -- yielding an upper limit to the net source count rate of
about 4.5 counts in 5000 seconds -- about half the upper limit in 2007
with Swift.  In this case, it is clear that the source must have
either faded or changed spectrum significantly since the deep Chandra
and XMM observations taken from 2002 through 2004.

\section{Discussion}
Three possibilities have been laid out for the nature of
XMMU~122939.7+075333.  One of these possibilities is that the source
is a red giant-black hole binary, with the change in brightness in the
2004 XMM observation caused by a grazing eclipse of the inner
accretion disk by a puffy, precessing outer disk (Shih et al. 2008);
this possibility is no longer viable because the large ratio of [O
III] to Balmer emission strongly favors an evolved donor star.
Alternatively, the accreting object may a stellar mass black hole
accreting from a WD in a short period binary system, or it may be the
result of a tidal detonation of a WD by an intermediate mass BH (Irwin
et al. 2010).  We can then consider the implications of the X-ray
variability on these different classes of models.

\subsection{WD-BH binary and triple models}
The strong variability seen from XMMU~122939.7+075333 is easy to
explain in a model where the accretion is taking place in a
hierarchical triple star system, with the inner binary being WD-BH
X-ray binary.  Such a scenario is the preferred means for forming a
WD-BH X-ray binary according to the theoretical work of Ivanova et
al. (2010).

An aspect of the triple star system that was not explored by Ivanova
et al. (2010) may have profound consequences for the observability of
the system.  The Kozai cycles that are invoked for grinding down the
system to Roche lobe overflow should continue after the system has
come into Roche lobe overflow.  The eccentricity of the inner binary
will then continue to oscillate.  It has been shown previously that
even small eccentricities can produce large changes in mass accretion
rates (Hut \& Paczynski 1984) -- the mass transfer rate should
increase as the density of material at the Roche lobe radius.  To
first order, that should give a dependence as $exp(eR/h)$, where $e$
is the eccentricity of the binary, $h$ is the scale height of the
star, and $R$ is the radius of the star.  The value of $h/R$ is
typically $10^{-4}$ for a main sequence star, and should be a bit
smaller for WDs unless they are very hot.  A triple system
could then be expected to produce an accretion rate which is modulated
substantially on the timescale for which the eccentricity changes due
to the Kozai cycles, even for very small eccentricity changes.

A convenient form for the period of eccentricity variations:
\begin{equation}
P_e = P_1 \left(\frac{m_0+m_1}{m_2}\right)\left(\frac{a_2}{a_1}\right)^3(1-e_2^2)^\frac{3}{2},
\end{equation}
where $P_1$ is the period of the inner binary, $m_2$ is the mass of
the outer star, and $a_2$ is the orbital separation of the outer star
from the center of mass of the inner binary is given by Ford et
al. (2000) -- see also Mazeh \& Shaham (1979).  In the case that
$m_2<<(m_0+m_1)$, we can use Kepler's third law to solve for $P_2$,
the period of the outer star in the hierarchical triple:
\begin{equation}
P_2= \sqrt{P_e P_1 \left(\frac{m_2}{m_0+m_1}\right) (1-e^2)^{-\frac{3}{2}}},
\end{equation}
with a correction term of $\sqrt{\frac{m_0+m_1+m_2}{m_0+m_1}}$ in the
case that $m_2$ is large enough for this term to be important.  The
accretion rate for XMMU~122939.7+075333 seems to have been roughly
constant from 1992 to 2004, but to have changed substantially between
2004 and 2010.  Taking illustrative values $m_0 \approx 10 M_\odot$,
$m_1=m_2=0.1 M_\odot$, $P_e=$40 years, and $P_1=$10 minutes, gives
$P_2$ of 60 days, a reasonable value for the period of the outer star
in a hierarchical triple formed dynamically in a globular cluster, as
the binary will still be ``hard.''

We note that the stellar mass BH scenarios for XMMU~122939.7+075333
require a mildly super-Eddington accretion rate.  The luminosity of
XMMU~122939.7+075333 in its bright states is above the Eddington
luminosity for a stellar mass BH, and the spectral shape of the source
is considerably softer than those for stellar mass BHs.  These
properties of the source can all be reconciled with the idea that this
source, in its bright states is in the ``ultraluminous state''
(Gladstone et al. 2009 -- see also Soria et al. 2007 for a somewhat
different scenario producing similar observables), in which a cool
photosphere develops in the inner accretion flow due to radiation
pressure.  One would also expect strong disk winds from such radiation
pressure dominated systems (e.g. Oosterbroek et al. 1997; Blundell et
al. 2001).  This gives good reason to believe that the $\dot{m}$ may
be changing substantially over the 10 year bright phase from
1994-2004, even if the luminosity changes rather little -- the X-ray
luminosity in the ultraluminous state is likely to change much more
slowly than linearly with mass transfer rate.  In the context of this
scenario, the disk wind may also be responsible for obscurring the
inner accretion disk, with variations in the disk wind causing events
like the decrease in brightness seen in the 2004 XMM observation.

\subsection{Tidal destruction scenarios}
In scenarios where the broad [O~III] emission lines arise from
outflows generated in a tidal disruption or detonation of a white
dwarf, {\it ad hoc} scenarios must be invoked for strong variability
of the X-ray source.  Models of the mass flow of tidal disruption
debris onto the disruptor have been calculated for the case of
supermassive BHs disrupting main sequence and giant stars.  They find
$\dot{m} \propto t^{-5/3}$ from analytic calculations (Rees 1988) and
from smoothed particle hydrodynamics calculations (Bogdanovi\'c et
al. 2004), with slightly shallower relations when accretion is
modulated by a disk (Cannizzo et al. 1990).

A generic feature of the solutions is that the luminosity decline is
not terribly steep.  Given that the luminosity of XMMU~122939.7+075333
varied by factors of a few, at most, in the observations collected
from the 1992 ROSAT observations through the 2004 XMM observations, a
variation of a factor of 20 or so from 2004 to 2010 should not be
expected.  Since the time baseline is longer between the ROSAT
observations and the long XMM observation than from the long XMM
observation to the newest {\it Chandra} observation, the fractional
change in the X-ray luminosity expected from a tidal disruption event
is expected to be smaller in the second period of time than the first.
Additionally, the source was not detected with Einstein. The upper
limits from Einstein are about $\sim10^{40}$ ergs/sec (Harris et
al. 1993) -- this would also be a problem for a tidal disruption
model.  

It has additionally been suggested that the optical line emission from
XMMU~122939.7+075333 might have been produced through the tidal
detonation of a WD, rather than a tidal destruction (Irwin et
al. 2010), a scenario previously modelled by Rosswog et al. (2009).
Tidal detonation requires the production of enough iron that one would
expect to see iron lines in the optical spectra unless the explosion
products were inhomogeneous in a fine-tuned manner.  Additionally,
Rosswog et al (2009) suggest that the only difference between the long
term X-ray variability of a detonation and a disruption event is that
the total mass reservoir in a detonation event is likely to be a
factor of $\sim3$ smaller; the problems in explaining the long-term
lightcurve of the source thus remain.

The strong X-ray variability we report in this paper therefore cannot
be due to the secular evolution of an accreting intermediate mass
BH system which has tidally disrupted a WD.  In IMBH
scenarios, the accretion will be at well below the Eddington rate.  As
a result, strong disk winds are unlikely to develop.  The short
timescale variability seen on 1 January 2004 is thus difficult to
explain, and it is also then difficult to explain the longer timescale
variability through changes in foreground absorption.

\subsection{Pure absorption changes}
We can also consider the effects of changing only the foreground $N_H$
without changing the intrinsic luminosity of the source.  Oosterbroek
et al. (1997) observed the 1989 outburst of V404 Cyg and found that
its spectral variability was well fitted by including a variable $N_H$
which ranged from $5\times10^{22}$ to $1.6\times10^{23}$ cm$^{-2}$
near the peak of the source's outburst, when it is likely to have been
in a super-Eddington state.  It then seems reasonable that a change in
$N_H$ of $\sim10^{23}$ cm$^{-2}$ is possible for very bright X-ray
sources.  Using W3PIMMS, we compute the count rate expected for this
source if it is modelled by a 0.2 keV blackbody with an intrinsic
luminosity of $4\times10^{39}$ ergs/sec with a foreground absorption
of $10^{23}$ cm$^{-2}$, and find that the expected count rate on
ACIS-S is about $1.5\times10^{-5}$ counts/sec -- a factor of about 10
below our detection limit.  It is thus possible that the source has
not varied intrinsically, and that just the foreground absorption has
varied.  The non-detection with Swift in March of 2010 argues (albeit
only at the $\sim$ 3$\sigma$ level) against a change in absorption as
the reason for the fading seen in the Chandra observations, unless
that change is rather long lived.  It should be noted that such a
long-lived change might be expected for the case for a disk wind from
a precessing disk (if e.g. the inclination angle of the disk wind
changed from one from not intercepting the line of sight to one
intercepting it), and that the one Galactic X-ray binary with a
strong, persistent disk wind, SS~433 (Blundell et al. 2001) is well
known to have a strong precession (e.g. Margon 1984).

If the X-ray variability is due to real changes in the central engine
luminosity, then the [O III] luminosity should respond to those
changes on a light travel time; if it is due to changes in the
absorption, the response of [O III] should be much weaker, since most
of the photoionizing emission is directed along lines of sight other
than the one between the source and the Earth.  An approved {\it
Chandra} + {\it Gemini} campaign will provide the first test of
whether the X-ray source is still on, and whether the [O III]
luminosity has responded to any changes in X-ray luminosity.

We note that we have done these calculations assuming that the
absorbing material has the cosmic abundances of Morrison \& McCammon
(1983), which are embedded within W3PIMMS, rather than the nearly
hydrogen-free abundances expected on the basis of the optical spectra
of RZ~2109.  Since the extinction in the soft X-rays is predominantly
due to carbon and oxygen, it is not intrinsically problematic that
there is no hydrogen present.  The column densities of carbon and
oxygen will be nearly the same in a hydrogen-free absorber as they
would be in a solar composition absorber with the fitted $N_H$.

\section{Conclusions}
We have reported large amplitude X-ray variability from
XMMU~122939.7+075333, the first strong candidate for being a BH X-ray
binary in a globular cluster.  We have shown that the combination of
X-ray luminosity, X-ray variability on both short and long timescales,
and optical line emission are all consistent with the idea that this
system is, rather than being an X-ray binary, a hierarchical triple
system with its inner binary composed of a WD and a stellar mass BH.
Better sampling of the long timescale X-ray variability of the system
is needed to test this idea more thoroughly.  We have also shown that
in the proposed scenario in which the system has an intermediate mass
BH accretor has serious problems reproducing both the short and long
term variability.

\section{Acknowledgments}
TJM thanks the European Union for support under FP7 grant 215212:
Black Hole Universe.  AK thanks NASA for support under Chandra grant
GO0-11111A and HST archival program HST-AR-11264.  SEZ thanks NASA for
support under grants NNX08AJ60G and Chandra GO0-11105X.  This research
is supported in part by an NSF Faculty Early Career Development
(CAREER) award (AST-0847109) to KLR.  We thank the Swift team for
awarding target of opportunity observations to look at
XMMU~122939.7+075333.  TJM thanks Cole Miller and Ian McHardy for
useful discussions.  We thank an anonymous referee for a brief report
with a few very useful suggestions that have hopefully improved the
clarity of the paper.

\label{lastpage}

\end{document}